# Cluster Expansion of Electronic Excitations: Application to fcc Ni-Al Alloys

H. Y. Geng,[1,2] M. H. F. Sluiter,[3] and N. X. Chen[1,4]

[1]*Department of Physics, Tsinghua University, Beijing 100084, China*

[2]*Laboratory for Shock Wave and Detonation Physics Research,*
*Southwest Institute of Fluid Physics,*
*P. O. Box 919-102, Mianyang Sichuan 621900, China*

[3]*Institute for Materials Research, Tohoku University, Sendai, 980-8577 Japan*

[4]*Institute for Applied Physics, University of Science and Technology, Beijing 100083, China*

## Abstract

The cluster expansion method is applied to electronic excitations and a set of effective cluster density of states (ECDOS) are defined, analogous to effective cluster interactions (ECI). The ECDOS are used to generate alloy thermodynamic properties as well as equation of state (EOS) of electronic excitations for the fcc Ni-Al systems. When parent clusters with small size, the convergence of the expansion is not so good but the electronic density of state (DOS) is well reproduced. However, the integrals of the DOS such as the cluster expanded free energy, entropy and internal energy associated with electronic excitations are well described at the level of the tetrahedron-octahedron cluster approximation, indicating the ECDOS is applicable to produce electronic ECI for cluster variation method or Monte Carlo calculations. On the other hand, the Grüneisen parameter, calculated with first-principles methods, is not any longer a constant and implies that the whole DOS profile should be considered for EOS of electronic excitations, where ECDOS adapts very well for disordered alloys and solid solutions.

## I. INTRODUCTION

The first-principles theory of alloy based on density functional theory (DFT) has received much attention in recent years.[1,2] Until now, theoretical investigations have focused mainly on phase stability at ambient pressure.[2,3] However, the equation of state (EOS) of solid solutions and mixtures is essential for understanding phase stability at high pressures and high temperatures. Static or dynamic process of pressure loading may change the stable structures of alloys and associated physical properties[4,5] as segregation and order-disorder transformation can occur. To understand these phenomena deeply, knowledge about the corresponding EOS is necessary.

Although the EOS theory for pure substances is well developed,[6] the extension to alloys is crude. The cluster expansion method (CEM) might be useful in this field[7] because it is a natural generalization of the mixing model that is currently widely used for the EOS of alloys. With the CEM one can provide precise thermodynamic properties of alloys if chemical energies, vibrational free energies and electronic excitation free energies are available for a set of superstructures.[1,8] Here the contribution of electronic excitations is given special attention. Although the chemical and vibrational contributions have been investigated intensively in recent years,[1,9] the effect of the electronic excitations and the corresponding convergence of the CEM has received scant attention.[10] This is understandable because at ambient pressure the temperature range of interest for alloys is about $10^3$K and below, where electronic excitations are negligible. However, electronic excitations become important at high pressures and temperatures for its magnitude is in proportion to $T^2$ (in contrast to $\ln T$ for lattice vibrations) and becomes dominant at enough high $T$. This range of temperature is well interested for EOS theories and experiments, which partially motivates the present work in order to construct a complete EOS model for alloys.

In this paper we shall illustrate the convergence of the CEM of electronic excitations based on DFT calculations of the Ni-Al system with underlying fcc lattice. Conventional EOS models for electronic excitations, such as the free-electron approximation and the Thomas-Fermi theory,[6] though with the virtue of simplicity, are less accurate than modern DFT[11] and are difficult to generalize beyond the simplest level for alloys. We present the basic theoretical model of the CEM for electronic excitations and its contribution to the EOS. Next, first-principles calculations of a set of fcc Ni-Al superstructures are used to verify and analyze the convergence of the CEM. The electronic Grüneisen parameter is evaluated as a function of temperature, atomic volume and composition and the applicability of the Mie-Grüneisen EOS for electronic excitations is discussed.

## II. THEORETICAL MODEL

A convenient representation of an alloy system is the Ising model.[3] In the case of a binary alloy system a spinlike occupation variable $\sigma_s$ is assigned to each site $s$ of the parent lattice where $\sigma_s$ takes the value $-1$ or $+1$ depending on the type of atom occupying the site. A particular arrangement of spins of the parent lattice is called a configuration and can be represented by a vector $\sigma$ containing the value of the occupation variable for each site in the parent lattice. Then all the thermodynamic information of an alloy is contained in the partition function[1]

$$Z = \sum_L \sum_{\sigma \in L} \sum_{v \in \sigma} \sum_{e \in v} \exp\left[-\beta E\left(L, \sigma, v, e\right)\right], \qquad (1)$$

where $\beta = 1/(k_B T)$; $L, \sigma, v$ and $e$ specify the parent lattice, configuration, atomic displacement from the ideal lattice site and particular electronic state, respectively. Following the coarse graining process of Ceder,[12] Eq.(1) can be approximated as

$$Z(V, T) = \sum_\sigma \exp\left\{-\beta \left[E_s\left(\sigma, V\right) + F_v\left(\sigma, V, T\right) + F_e\left(\sigma, V, T\right)\right]\right\} \qquad (2)$$

for a specific parent lattice and atomic volume $V$, the free energy then becomes[8]

$$F(V, T) = \sum_\sigma \rho_\sigma(V, T) \left[E_s\left(\sigma, V\right) + F_v\left(\sigma, V, T\right) + F_e\left(\sigma, V, T\right)\right]$$
$$+ k_B T \sum_\sigma \rho_\sigma(V, T) \ln \rho_\sigma(V, T) \qquad (3)$$

where $\rho_\sigma(V, T)$ is the configurational density matrix defined as

$$\rho_\sigma(V, T) = \frac{\exp\left\{-\beta \left[E_s\left(\sigma, V\right) + F_v\left(\sigma, V, T\right) + F_e\left(\sigma, V, T\right)\right]\right\}}{Z(T, V)}. \qquad (4)$$

In Eq.(3), $E_s$, $F_v$ and $F_e$ denote the contributions arising from chemical interaction, lattice vibrations and electronic excitations at a given configuration, atomic volume $V$ and temperature $T$, respectively. The last term in Eq.(3) specifies the negative of configurational entropy times $T$. Here we are just interested in the contribution of electronic excitations and after dropping irrelevant terms, the remaining free energy term is

$$F\left([\rho], V, T\right) = \sum_\sigma \rho_\sigma(V, T) F_e\left(\sigma, V, T\right). \qquad (5)$$

Here we have defined the free energy as a functional of the configurational density matrix. The equilibrium free energy is then obtained by the variational principle by minimization of Eq.(5) with respect to $\rho$.

The remaining problem is to determine $F_e$ for a set of configurations. If we consider the whole crystal lattice containing all sites, then each $\sigma$ corresponds to a specific ordered structure whose $F_e$ can be calculated with first-pinciples methods directly and Eq.(5) is solved. Unfortunately, this scheme involves too many variables and is impractical. Therefore, an approximation is needed to evaluate Eq.(5) by restricting $\sigma$ to $\sigma_m$ of some largest clusters (the so-called parent clusters). In this case, $\sigma_m$ no longer relates to a specific ordered structure and special arrangements are required. Being different from the original proposal of CEM, we would like here to re-derive it again with another approach, which is more intuitive in physics, as follows: letting the configurational density $\rho_\sigma = 0$ if $\sigma$ cannot be reproduced by stacking (neither in overlap nor non-overlap manner) of some configurations $\sigma_m$ of the parent clusters, which means this structure cannot be described properly within such size of parent clusters and just drop it simply, otherwise relating these $\sigma_m$ to the ordered structure $\alpha$ characterized by $\sigma$ with a density of $\rho_{\alpha|\sigma_m}$, namely the probability of $\sigma_m$ configuration appears in $\alpha$ phase times the probability of $\alpha$ phase itself. where $\alpha|\sigma_m$ indicates the configuration of $\alpha$ phase is restricted to $\sigma_m$. Note this is not an one-to-one mapping (it is possible that there are many types of $\sigma_m$ which describe the same structure, vice versa), but is helpful to categorize $\sigma_m$ according to structure. An equivalent expression is $\cup \sigma \xrightarrow{\sigma|\sigma_m} \bigcup_\alpha \sigma_m^\alpha$ ($\sigma_m^\alpha$: $\sigma_m \in$ structure $\alpha$). Here the union means considering all possible configurations on the lattice and $\sigma_m \in \alpha$ indicates $\alpha$ structure can be reproduced by either overlap or non-overlap stacking of $\sigma_m$ (together with other configurations, if necessary). Then we have

$$\sum_\sigma \rho_\sigma F_e(\sigma) \to \sum_\alpha \sum_{\sigma_m \in \alpha} \rho_{\alpha|\sigma_m} F_e(\alpha|\sigma_m). \qquad (6)$$

In this equation we have omitted irrelevant variables and $F_e(\alpha|\sigma_m)$ is a mere formality, denotes the contribution to the electronic free energy of $\alpha$ phase from $\sigma_m$ configuration. Recall that the configuration density matrix can be expanded on an orthogonal and complete basis formed by correlation functions defined as $\xi_i^\alpha = \langle \sigma_i \rangle^\alpha$,[13] where the average operation is constrained on structure $\alpha$ and $i$ denotes cluster type, one may rewrite Eq.(6) as

$$\sum_\alpha \sum_{\sigma_m \in \alpha} \rho_{\alpha|\sigma_m} F_e(\alpha|\sigma_m) = \sum_\alpha \sum_i \left[ \sum_{\sigma_m \in \alpha} \mathbf{M}_{\alpha|\sigma_m,i} F_e(\alpha|\sigma_m) \right] \xi_i^\alpha = \sum_\alpha \sum_i \lambda_i^\alpha \xi_i^\alpha. \qquad (7)$$

In this way the correlation functions $\xi_i$ dependent free energy functional is given by

$$F([\xi_i], V, T) = \sum_i \left[ \sum_\alpha \lambda_i^\alpha (V, T) \xi_i^\alpha / \xi_i \right] \xi_i = \sum_i \lambda_i (V, T) \xi_i. \tag{8}$$

$\lambda_i (V, T)$ are the effective cluster interactions (ECI) and the convergence of the CEM is characterized as

$$\sum_\alpha \lambda_i^\alpha (V, T) \xi_i^\alpha / \xi_i \to \lambda_i (V, T) \tag{9}$$

for any available $\xi_i$ on parent lattice. From Eqs.(7,8) one gets, for a set of ordered structures, $F^\alpha (V, T) = \sum_i \lambda_i (V, T) \xi_i^\alpha$. Thus the ECI can be obtained approximately with

$$\lambda_i (V, T) = \sum_\alpha \left( \xi^{-1} \right)_i^\alpha F^\alpha (V, T), \tag{10}$$

which is the well-known Connolly-Williams method[14] and $F^\alpha$ is computable with standard DFT methods.

It is obvious now that the CEM rests on two assumptions: one is that the ECI corresponding to clusters larger than the parent clusters are negligible, see Eqs.(6) and (8), which implies that interactions are short-ranged; and the other is that limiting values of the ECI should be achieved with a small, finite set of ordered structures (Eq.(9)), which is not very clear before. The latter implies that the underlying lattice should be compact and highly symmetric which is the case for metallic alloys with fcc/bcc/hcp crystal structures, otherwise one cannot model the coordinate environments of different configurations properly by just a small set of typical structures. Actually, we have observed the convergent difficulty of CEM on the chemical energy of ThMn$_{12}$ structure, indicating it should be further improved before can be applied to rare earth materials.

For electronic excitations there are three different levels of approximation for the free energy: the first level is finite-temperature DFT where the temperature dependence of the Fermi-Dirac distribution, the Fermi energy, and the density of states (DOS) is included in the self-consistency loop; the second level assumes that the DOS has no explicit temperature dependence; and the third level is known as the Sommerfeld approximation[10] where the DOS is not only assumed to be temperature independent, but also to be constant with $\epsilon$ near the Fermi energy, so that the free energy is characterized by $n(\epsilon_F)$ only. It was found that the temperature dependence of the DOS is weak so that level 2 provides an accurate approximation as compared with level 1.[10] Therefore we adopted approximation level 2 for calculating the electronic excitation free energies of a set of ordered fcc superstructures.

The Sommerfeld approximation was found to be too crude and has been employed for the purpose of comparison only.

At approximation level 2, all thermodynamic quantities are determined by the $T=0$ DOS. In particular, the free energy is given by $F_e(V,T) = E_e(V,T) - TS_e(V,T)$ where

$$S_e(V,T) = -\int_{-\infty}^{+\infty} n(\epsilon,V)\{f(\epsilon,T)\ln f(\epsilon,T) + [1-f(\epsilon,T)]\ln[1-f(\epsilon,T)]\} d\epsilon \qquad (11)$$

and

$$E_e(V,T) = \int_{-\infty}^{+\infty} \epsilon n(\epsilon,V) f(\epsilon,T) d\epsilon - \int_{-\infty}^{\epsilon_F} \epsilon n(\epsilon,V) d\epsilon, \qquad (12)$$

where $f(\epsilon,T)$ is the Fermi-Dirac distribution function. Note that $E_e$ is defined such that it vanishes at $T=0$ for each structure because here we are interested in electronic excitations only. If we arbitrarily define $\epsilon_F=0$, then the configuration and volume dependences of the free energy is given by $n(\epsilon,V)$ solely and based on Eqs.(11) and (12) an operator $\hat{F}$ is defined as

$$F_e^\alpha(V,T) = \hat{F}[n^\alpha(\epsilon,V)]. \qquad (13)$$

The linearity of Eq.(10) permits one to rewrite it as

$$\lambda_i(V,T) = \hat{F}\left[\sum_\alpha (\xi^{-1})_i^\alpha n^\alpha(\epsilon,V)\right] = \hat{F}[d_i(\epsilon,V)],$$

where $d_i(\epsilon,V)$ is the effective cluster DOS (ECDOS), the analogue of ECI, defined by

$$d_i(\epsilon,V) = \sum_\alpha (\xi^{-1})_i^\alpha n^\alpha(\epsilon,V). \qquad (14)$$

Note here however, that the linearity of the electronic free energy in the DOS is only approximate based on level 2 at finite temperature. When $T$ is high enough and the electronic chemical potential changes distinctly, above equations will break down. Evidently, the convergence of the electronic excitation CEM is completely determined by the behavior of the ECDOS: how fast the ECDOS approach their limits by including more ordered structures to produce them and how fast they tend to zero with increased cluster size.

Similarly, the EOS is also determined by the DOS. For any energy $\epsilon$, there is a corresponding frequency $\omega$ which satisfies $\epsilon = \hbar\omega$. However, it is the electronic DOS $n(\epsilon,V)$ which describes the volume dependence of electronic excitation contribution and is in complete analogy with phonon frequencies in the case of lattice vibrations. In this way, following Grüneisen,[6] we can define the electronic Grüneisen parameter as

$$\Gamma_e(\epsilon,V) = -\frac{V}{n(\epsilon,V)}\frac{dn(\epsilon,V)}{dV}, \qquad (15)$$

which has the same physical implications as the lattice vibrational Grüneisen parameter. Considering that all thermodynamic properties are generated from the DOS, a more practical average of $\Gamma_e$ can be achieved by weighting with respect to $n(\epsilon, V)$,

$$\gamma_e(V, T) = \frac{\hat{F}[\Gamma_e(\epsilon, V) n(\epsilon, V)]}{\hat{F}[n(\epsilon, V)]}. \tag{16}$$

Using Eqs.(13), (15) and (16), it is easy to prove that the EOS for electronic excitations can be written in Grüneisen fashion

$$P_e(V, T) = \frac{\gamma_e(V, T)}{V} F_e(V, T). \tag{17}$$

It is necessary to point out that in the limit of the free electron approximation, $\gamma_e$ of Eq.(16) approaches to -2/3, the negative value of the conventional electronic Grüneisen parameter.[15] This is reasonable because within the free electrons approximation, $F_e$ approaches the negative of the electronic internal energy.

## III. CALCULATIONS AND DISCUSSIONS

The DOS of a set of fcc Ni-Al superstructures have been calculated with the generalized gradient approximation (GGA)[16] using the CASTEP (CAmbridge Serial Total Energy Package)[17,18] with fcc lattice parameter $a$ from 2.5 to 4.6Å with an interval of 0.1Å. The set contains the following ordered structures: fcc A and B, $L1_0$, $L1_2$ ($A_3B$ and $AB_3$), $DO_{22}$ ($A_3B$ and $AB_3$), $MoPt_2$ type of order ($A_2B$ and $AB_2$), $A_2B_2$ (phase 40 in Kanamori's notation), $L1_1$ and C2/m ($A_2B$).[19] The last structure is employed to verify the convergence of the electronic excitation CEM while other structures are used to derive ECI and ECDOS. The calculations are performed using ultrasoft pseudopotentials[20] with a cutoff kinetic energy for planewaves of 540 eV. Integrations in reciprocal space are performed in the first Brillouin zone with a grid with a maximal interval of 0.03Å$^{-1}$ generated with the Monkhorst-Pack[21] scheme. The energy tolerance for self-consistency convergence is 2$\mu$eV/atom for all calculations. Note that all physical quantities in this section are given for per atom.

### A. Convergence of electronic excitation CEM

As mentioned in the previous section, the cluster expansion of electronic excitations is essentially the expansion of the DOS. All thermodynamic quantities are then generated from the free energy (or partition function) depending on the ECDOS and correlation functions

$$F\left(\left[\xi_i\right], V, T\right) = \sum_i \hat{F}\left[d_i\left(\epsilon, V\right)\right] \xi_i. \tag{18}$$

Here the ECDOS are derived with Eq.(14) using the tetrahedron-octahedron (T-O) approximation with eleven structures. The corresponding contributions from the null, point, tetrahedron and octahedron clusters at fcc lattice parameter $a$=3Å are illustrated in figure 1. We see that the contribution of the point is mostly negative because the DOS of Al is much lower than that of Ni. With this size of parent clusters, the convergence of the ECDOS is not so good. The contribution from the largest cluster, the octahedron, is still considerable near the Fermi energy, implying that some correction may be introduced if still larger clusters are involved. Note that far away from the Fermi energy, octahedron fluctuations are suppressed and the CEM converges.

The DOS of any stucture can be obtained with the ECDOS. The C2/m (Al$_2$Ni) structure with $a$=3Å, which is excluded when deriving the ECDOS, is employed to check the capability of ECDOS to predict the DOS as is shown in figure 2. Except for some detailed features, the main profile of the *ab initio* DOS is reproduced.[22] The discrepancy at the high energy side is unimportant because this range relates to unoccupied bands which are not sampled by the CEM. In this sense, the convergence of the CEM for the DOS is better than expected on the basis of figure 1.

For partial or complete disordered state, it is difficult to calculate the DOS directly with DFT methods, and ECDOS provides an effective shortcut for this purpose. The DOS generated from ECDOS of a disordered structure with composition Al$_2$Ni is shown in figure 3 for comparison. Although the precision of current case is not high enough because of the employed largest cluster contains only six points, the reproduced DOS can always be greatly improved when much larger parent clusters and more superstructures are involved. Figs. 2 and 3 show that the disordering process has significant influence on the DOS around 2.5eV below the Fermi energy. It appears that the CEM provides an efficient way to examine the effect of partial or complete disorder.

Since the ECDOS cannot reproduce the DOS exactly with small parent clusters, the quality of corresponding cluster expanded thermodynamic quantities depends on temperature. Fortunately, these quantities are all integrations of the DOS and become less sensitive to its detailed structure. Therefore a better convergence of the CEM on these quantities is ex-pected. Figure 4 shows the cluster contributions to the internal energy, free energy and the entropy term, respectively. The cluster labels correspond to the null, the point, the nearest-neighbor (NN) pair, the next-nearest-neighbor (NNN) pair, the equilateral NN triangle, the isosceles triangle formed by one NNN and two NN pairs, the equilateral NN tetrahedron, the irregular tetrahedron consisting of one NNN and five NN pairs, the square formed by four NN pairs, the pyramid, and the octahedron, respectively. We find the convergence of the CEM for the energetic quantities at the T-O level is quite good. Of course, if larger clusters are involved, a small correction is still expected for the third NN pair and other related clusters which might have larger contribution than octahedron are not considered here. The electronic excitation free energy of C2/m ($Al_2Ni$) at $a=3$Å as computed *ab initio* and as obtained from the integrated ECDOS and as obtained from the Sommerfeld approximation are shown as functions of temperature in figure 5. It illustrates that the limiting values of the ECDOS (see Eq.(9)) are almost reached with just eleven structures. However the quality of convergence is somewhat reduced at high temperatures because the limiting error is magnified by a factor of $T^2$.

### B. Electronic Grüneisen parameter

The Mie-Grüneisen EOS is very efficient when applied to lattice vibrations.[6,15] It can be generalized to the case of electronic excitations in alloys with Eqs.(16-17). Here we examine closely the variation of the Grüneisen parameter in a coordinate space consisting of $T$, $a$ (relates to atomic volume $V=a^3/4$) and aluminum concentration $c_{Al}$. The merits of the Mie-Grüneisen EOS are based on the fact that the Grüneisen parameter is rather constant for a wide range of $T$ and $V$, resulting in simplification and reduction of calculations. Therefore it is necessary to check whether this advantage still holds for electronic excitations. If not, what is the most convenient approach for electronic EOS.

Using the ECDOS, the electronic Grüneisen parameter is calculated according to its definition Eq.(16). Figure 6 shows its variation as a function of $T$ and $a$ in $L1_2$ $Ni_3Al$ phase. The wrinkles on the surface are due to the limited precision in the calculations which is exacerbated by the fact that $\gamma_e$ is a derivative (although smooth pressure and free energy surfaces can be obtained in present precision, more accurate calculations are needed for smoother $\gamma_e$, which is in proportion to the ratio of pressure and free energy and bearing higher singularity). It is more evident at low temperatures as shown in figure 7. We may conclude from figure 6 that $-\gamma_e$ has different behaviors along $a$ at different fixed $T$. At low $T$ it shows various trends with $a$ (see figure 7), but at high $T$ is always decreased with increasing $a$. For very small $a$ and a range of $T > 10^4$K, $\gamma_e$ becomes almost $T$ independent.

In complete disordered Ni-Al, the dependence of $\gamma_e$ on composition is readily calculated without the need to employ the cluster variation method as is demonstrated in figure 8. It is interesting to point out that the Grüneisen parameters of transition and non-transition metals (here nickel and aluminum) have opposite variations with $a$. With increasing $a$, $-\gamma_e$ of nickel always decreases while that of aluminum increases. This difference between transition and non-transition metals has not been noticed before and is believed to be directly related to the nature of the $d$ electrons. Here also, it is difficult to find a range of $a$ or $c_{Al}$ where $\gamma_e$ remains constant.

Figs. 6 and 8 show that the free-electron approximation breaks down completely. The value of $\gamma_e$ is quite different from that of the free-electron approximation. Moreover, it is almost impossible to express $\gamma_e$ in a simple analytical form, which seems a little contrary to the original intention of Mie-Grüneisen EOS and diminishes its usefulness. It appears that generally speaking the Mie-Grüneisen EOS for electronic excitations is just a mere formality and Eq.(16) must be computed accurately for a reasonable EOS model, which reveals the necessity to implement ECDOS for alloys and solid solutions under high-temperatures and pressures, especially for disordered states.

## IV. CONCLUSION

In summary, we show in this paper that the cluster expansion of electronic excitations is determined completely by the corresponding DOS. A set of ECDOS analogous to ECI is defined and all thermodynamic properties of alloys relating to electronic excitations can be reproduced from it. When the size of used parent clusters is small, the convergence of ECDOS is not so good, while the reproduced DOS is acceptable by and large. As the physical quantities of interest for EOS theory (e.g., free energy, internal energy or entropy) are all integrations over the DOS, the cluster expansion of these quantities is less sensitive to the detailed structure of DOS and the convergence is rather good at the level of the T-O approximation. In this sense, ECDOS is rather practical and applicable for producing ECI for CVM calculations or Monte Carlo simulations. The electronic Grüneisen parameter has been derived and expressed in terms of the ECDOS. It is shown to vary considerably as a function of lattice parameter, temperature and composition which implies Eq.(16) should be treated exactly for feasible electronic excitations equation of state theory and ECDOS is the most expedient and effective approach available so far to calculate it for disordered alloys and solid solutions.

Acknowledgments

This work was supported by the National Advanced Materials Committee of China. And the authors gratefully acknowledge the financial support from 973 Project in China under Grant No. G2000067101. Part of this work was performed under the inter-university cooperative research program of the Laboratory for Advanced Materials, Institute for Materials Research, Tohoku University.

---

[1] A. van de Walle and G. Ceder, Rev. Mod. Phys. **74**, 11 (2002).

[2] M. H. F. Sluiter, Y. Watanabe, D. de Fontaine, and Y. Kawazoe, Phys. Rev. B **53**, 6137 (1996).

[3] F. Ducastelle, *Order and Phase Stability in Alloys* (Elsevier Science, New York, 1991).

[4] A. Lindbaum, E. Gratz, and S. Heathman, Phys. Rev. B **65**, 134114 (2002).

[5] A. Alavi, A. Y. Lozovoi, and M. W. Finnis, Phys. Rev. Lett. **83**, 979 (1999).

[6] S. Eliezer, A. Ghatak, and H. Hora, *An Introduction to Equation of State: Theory and Applications* (Cambridge University Press, Cambridge, 1986).

[7] H. Y. Geng, N. X. Chen, and M. H. F. Sluiter, Phys. Rev. B **70**, 094203 (2004); Phys. Rev. B **71**, 012105 (2005).

[8] C. Colinet and A. Pasturel, J. Alloys Compd. **296**, 6 (2000).

[9] M. H. F. Sluiter, M. Weinert, and Y. Kawazoe, Phys. Rev. B **59**, 4100 (1999).

[10] C. Wolverton and A. Zunger, Phys. Rev. B **52**, 8813 (1995).

[11] J. Hafner, Acta. Mater. **48**, 71 (2000).

[12] G. Ceder, Comput. Mater. Sci. **1**, 144 (1993).

[13] D. de Fontaine, in *Solid State Physics*, ed. by H. Ehrenreich and D. Turnbull (Academic Press,

New York, 1994), vol. **47**, pp. 84.

[14] J. W. D. Connolly and A. R. Williams, Phys. Rev. B **27**, 5169 (1983).

[15] X. Xu and W. Zhang, *Theoretical Introduction to Equation of State* (Science Press, Beijing, 1986) (in Chinese).

[16] J. P. Perdew, K. Burke, and M. Ernzerhof, Phys. Rev. Lett. **77**, 3865 (1996).

[17] Accelrys Inc., CASTEP Users Guide (San Diego: Accelrys Inc., 2001).

[18] V. Milman, B. Winkler, J. A. White, C. J. Pickard, M. C. Payne, E. V. Akhmatskaya, and R. H. Nobes, Int. J. Quant. Chem. **77**, 895 (2000).

[19] M. Sluiter and P. E. A. Turchi, Phys. Rev. B **40**, 11215 (1989).

[20] D. Vanderbilt, Phys. Rev. B **41**, 7892 (1990).

[21] H. J. Monkhorst and J. D. Pack, Phys. Rev. B **13**, 5188 (1976).

[22] The prediction error at Fermi energy implies this set of ECDOS is not suitable for practical electronic analysis. However, as the convergence of ECDOS can be continuously improved with increasing parent clusters size and involving more and more superstructures, reproducing much more accurate DOS is expected. But since some quantum information has been integrated out when deriving ECDOS, it is inappropriate to analyse detailed electronic behaviors of atoms by ECDOS directly. The merits of ECDOS mainly lie on its capability to generate well-behaved electronic free energy and the corresponding ECI. In fact, the widely used chemical energy ECI is also determined by ECDOS completely, namely, $v_i(V) = \int_{-\infty}^{\epsilon_F} \epsilon d_i(\epsilon, V) d\epsilon$, which demonstrates that a little convergent error at quantum level cannot pollute the convergence of CEM on thermodynamics level. Analogously, CEM on lattice vibrational free energy is actually on the phonon DOS, and a set of phonon ECDOS can also be derived to generate the vibrational ECI.

**Figure Captions:**

FIG. 1: (Color online) Some effective cluster DOS for the fcc Ni-Al system. The fluctuations of tetrahedron and octahedron clusters near the Fermi energy imply the limited convergence.

FIG. 2: (Color online) DOS of C2/m structure as computed *ab initio* (solid) and as reproduced from the ECDOS (dash-dot). The occupied part of the profile is reproduced except for the underestimated valley near -5eV.

FIG. 3: (Color online) The DOS of disordered phase as obtained from the ECDOS (solid). Notice the peaks and valley near -2.5eV. Parts of ECDOS (dash-dot) copied from figure 1 are also presented for comparison.

FIG. 4: Cluster contributions to free energy $F$, internal energy $E$ and entropy term $TS$ at $T=10^4$K and $a=3$Å. The cluster labels are described in the text.

FIG. 5: The electronic excitation free energy as a function of $T$ for C2/m (Al$_2$Ni) computed by *ab initio* method comparing with those obtained from the integrated ECDOS and from the Sommerfeld approximation, respectively.

FIG. 6: (Color online) Negative of the electronic Grüneisen parameter in the coordinate space of $T$ and $a$. Contours are projected onto the bottom plane.

FIG. 7: Negative of the electronic Grüneisen parameter as functions of $a$ at low temperatures. See figure 6 for further reference.

FIG. 8: (Color online) Negative of the electronic Grüneisen parameter for the disorder Ni-Al alloys in the coordinate space of $c_{Al}$ and $a$ at $T = 3.5 \times 10^4$K. Contours are projected onto the bottom plane.

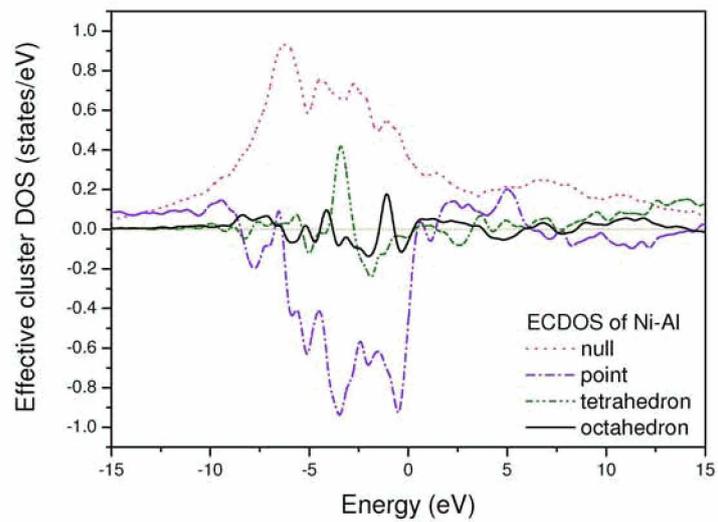

Figure 1, Geng et al, Journal of Chemical Physics

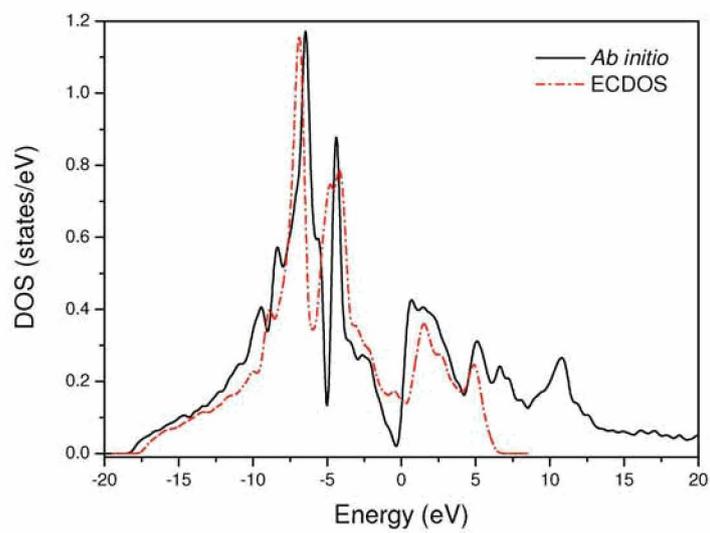

Figure 2, Geng et al, Journal of Chemical Physics

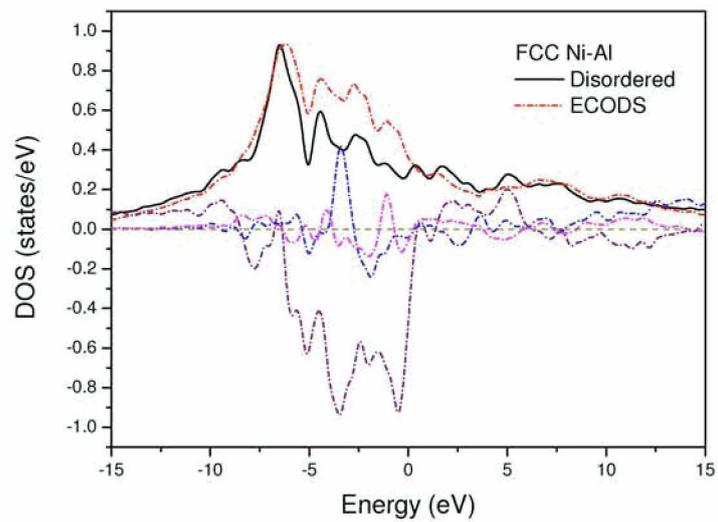

Figure 3, Geng et al, Journal of Chemical Physics

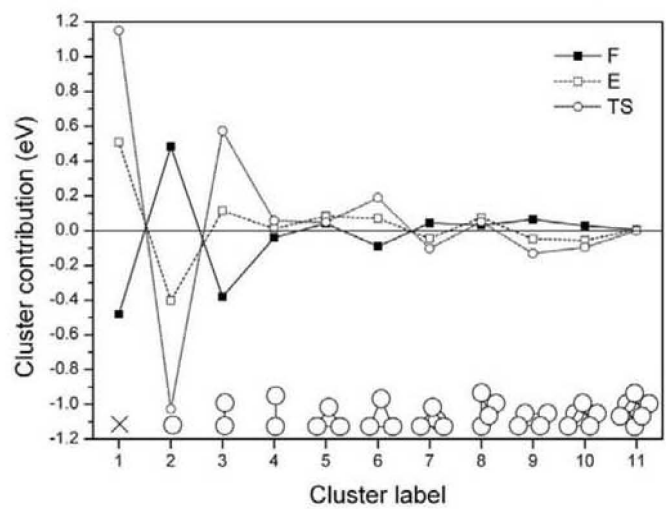

Figure 4, Geng et al, Journal of Chemical Physics

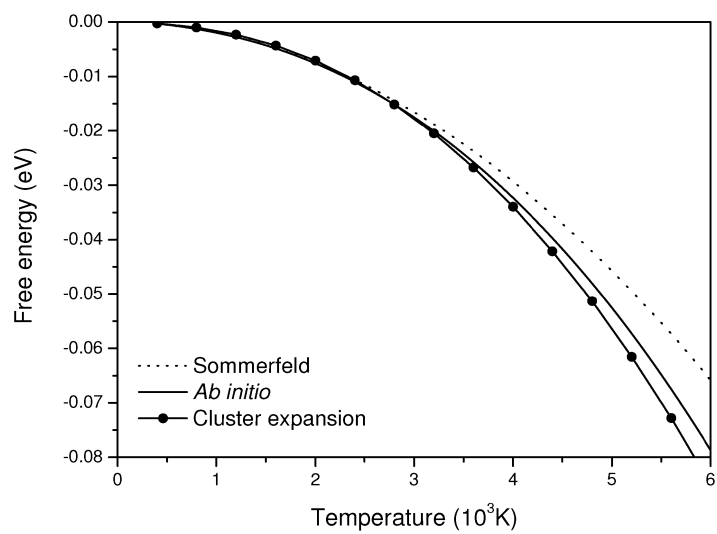

Figure 5, Geng et al, Journal of Chemical Physics

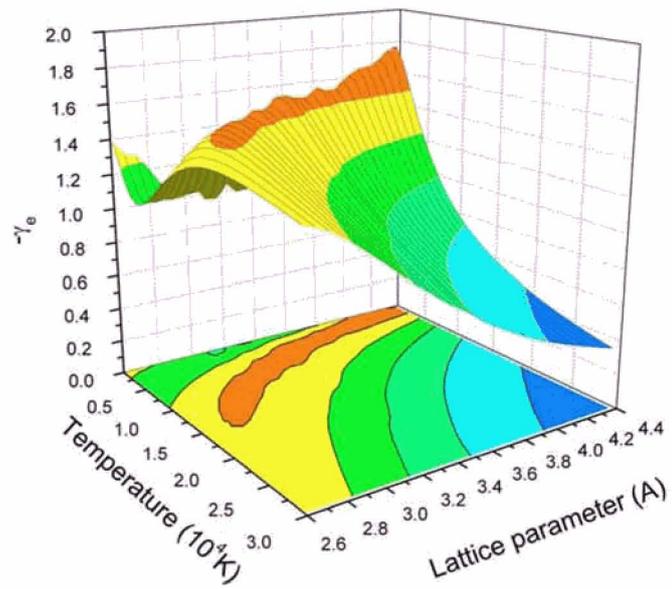

Figure 6, Geng et al, Journal of Chemical Physics

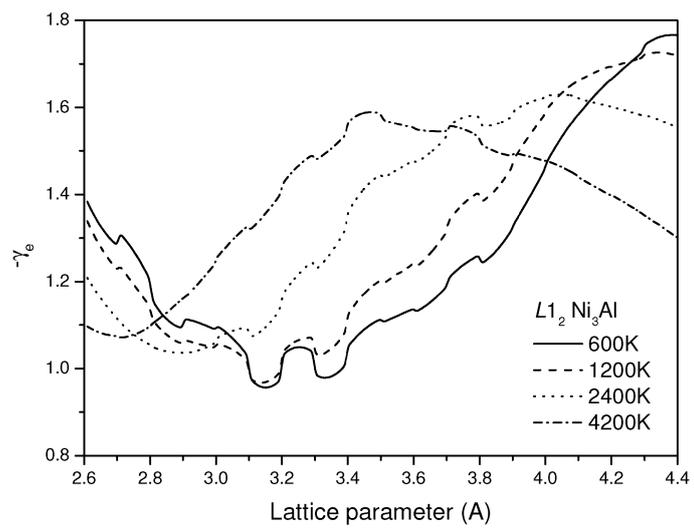

Figure 7, Geng et al, Journal of Chemical Physics

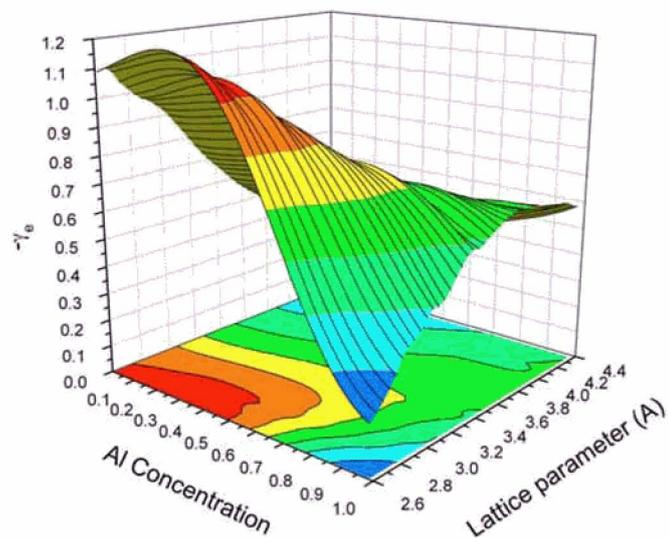

Figure 8, Geng et al, Journal of Chemical Physics